\documentclass[pdflatex,sn-mathphys-num]{sn-jnl}

\usepackage{graphicx}%
\usepackage{multirow}%
\usepackage{amsmath,amssymb,amsfonts}%
\usepackage{amsthm}%
\usepackage{mathrsfs}%
\usepackage[title]{appendix}%
\usepackage{xcolor}%
\usepackage{textcomp}%
\usepackage{manyfoot}%
\usepackage{booktabs}%
\usepackage{algorithm}%
\usepackage{algorithmicx}%
\usepackage{algpseudocode}%
\usepackage{listings}%
\newcommand{\REV}{\textcolor{black}}         

\begin{document}

\title[Article Title]{
Pyramidal charged domain walls in ferroelectric BiFeO$_3$}


\author*[1,2]{\fnm{Pavel} \sur{Marton}}\email{marton@fzu.cz}

\author[2]{\fnm{Marek} \sur{Pa\'{s}ciak}}

\author[2]{\fnm{Mauro} \fnm{A.P.} \sur{Gon\c{c}alves}}

\author[1]{\fnm{Ond\v{r}ej} \sur{Nov\'{a}k}}



\author[2]{\fnm{Ji\v{r}\'{i}} \sur{Hlinka}}

\author[3]{\fnm{Richard} \sur{Beanland}}

\author[3]{\fnm{Marin} \sur{Alexe}}


\affil[1]{\orgdiv{Institute of Mechatronics and Computer Engineering}, \orgname{Technical University of Liberec}, \orgaddress{\street{Studentská 2}, \city{Liberec}, \postcode{46117},  \country{Czech Republic}}}

\affil[2]{\orgdiv{Department of Dielectrics}, \orgname{Institute of Physics of the Czech Academy of Sciences}, \orgaddress{\street{Na Slovance 2}, \city{Praha}, \postcode{182 00}, \country{Czech Republic}}}

\affil[3]{\orgdiv{Department of Physics}, \orgname{University of Warwick}, \orgaddress{\street{Coventry CV4 7AL}, \postcode{10587},  \country{United Kingdom}}}


\abstract{
%
%
%
%
%
%
%

Domain structures play a crucial role in the electric, mechanical and other properties
of ferroelectric materials.
In this study, we uncover the physical origins of the enigmatic zigzag domain structure in the prototypical multiferroic material BiFeO$_3$.
Using phase-field simulations within the Landau-Ginzburg-Devonshire framework, we demonstrate that
spatially-homogeneous
defect charges result in domain structures that closely resemble those observed experimentally.
The acquired understanding of the underlying physics of pyramidal-domain formation may enable the engineering of new materials with self-assembled domain structures exhibiting defined domain periodicity at the nanometre scale, opening avenues for advanced applications.
}

\keywords{Ferroelectric BiFeO$_3$, Pyramidal domain walls, Zigzag domain walls, Charged domain walls, Transmission electron microscopy, Phase-field simulations}



\maketitle

\section{Introduction}\label{sec_introduction}

\begin{figure}[h]
\centering
\includegraphics[width=1.0\textwidth]{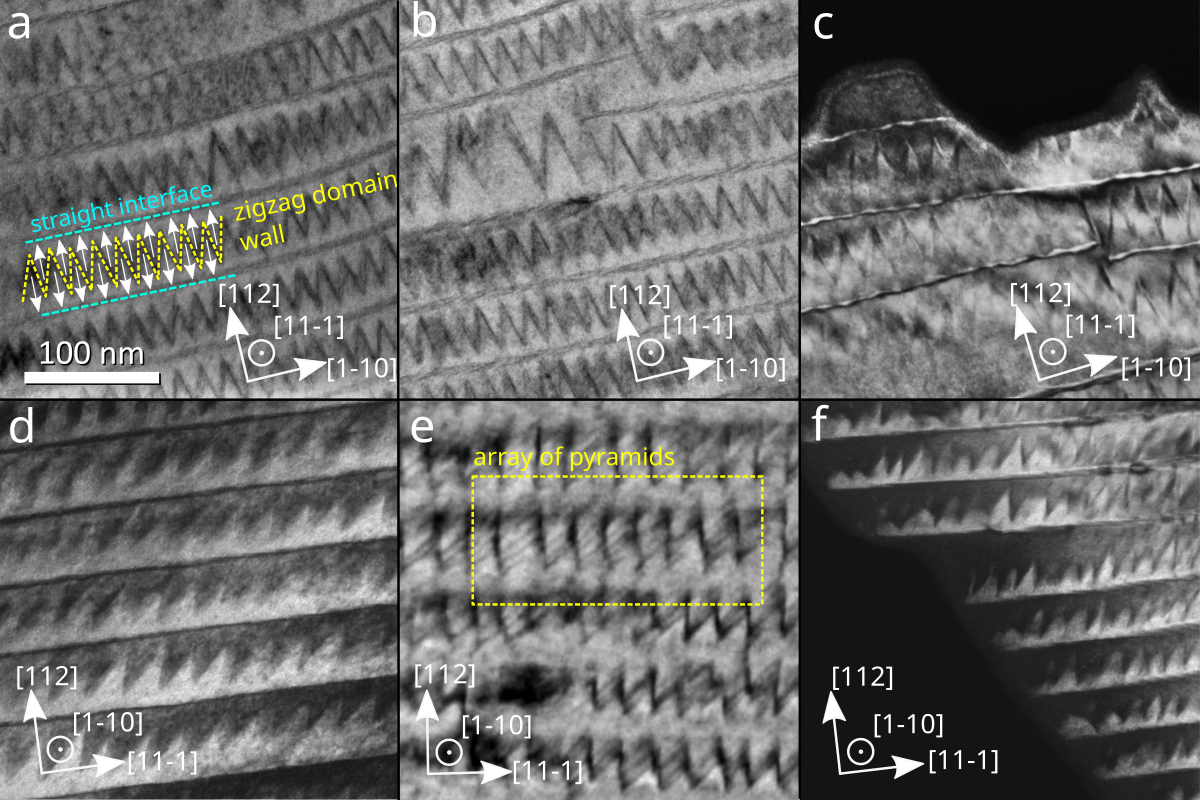}
\caption{
Pyramidal domain structures as observed using transmission electron microscopy viewed along two different directions. Upper row: incident beam direction close to $[11\bar{1}]$. Bottom row: incident beam direction close to $[1\bar{1}0]$.
a) Typical regular domain structure viewed along $[11\bar{1}]$ with symmetric triangles. The straight and zigzag interfaces form head-to-head and tail-to-tail 180$^\circ$ domain walls, respectively. Orientation of ferroelectric polarization between them is indicated by arrows. 
b) Imperfection in stacking of the straight head-to-head domain walls leads to triangles of various sizes. 
c) Domain structure in the thinnest part of the specimen.
d) Typical regular domain structure viewed along $[1\bar{1}0]$, with asymmetric triangles.
e) Image collected with a tilted specimen. The straight interfaces are not seen edge-on and are almost invisible, and the triangular domains can be seen to form an of array of pyramids.
f) Domain structure in the thinnest part of the sample.
Panels (c) and (f) demonstrate the stability of the domain structure in extremely thin specimens ($<20$ nm), close to the edge of the wedge-shaped FIB-cut lamella.
The indicated scale-bar in (a) is valid for all figures. Coordinate axes are approximate. 
%
%
%
}
\label{fig_TEM}
\end{figure}

Ferroic materials naturally evolve into multidomain states when undergoing the symmetry-breaking transition from their highest symmetry phases. The resulting domain configuration is dictated by multiple factors, including mechanical stress, electrical boundary conditions, impurities, growth process, history of the crystal, and others. Such multidomain materials can host phenomena very distinct from monodomain ones, both at the level of individual domain walls (separating domains) and/or as a consequence of particular domain structuring. The walls can be considered as 2D topological defects and as such in various materials have brought the addition of chirality~\cite{art_cherifi_2017,art_chauleau_2020}, polarity~\cite{art_aert_2012,art_an_2023}, conductivity~\cite{art_seidel_2009,art_beccard_2022}, electric susceptibility enhancement~\cite{art_xu_2014}, to mention just a few emergent properties. Similarly, at the mesoscale, a dense domain structure can significantly alter bulk material properties, and consequently, domain engineering has become a well explored path for designing advanced materials~\cite{art_wada_1999, art_nataf_2020,art_qiu_2020}. This has been accompanied by progress in understanding the processes that govern the development of domain structures. Owing to the mesoscale nature of the problem, computational phenomenological methods based on the Landau-Ginzburg-Devonshire theory have been particularly suited for simulating domain structures and their evolution~\cite{art_marton_2010, boo_ondrejkovic_2020}.


%
BiFeO$_3$ (BFO) is a prototypical multiferroic material with potential applications in spintronics, nonvolatile memory devices, sensors, actuators~\cite{art_wang_2020} and novel smart devices involving its ferroelastic, ferroelectric and antiferromagnetic orders which are present simultaneously~\cite{art_chaudron_2024}.
It is rhombohedral at room temperature, with large spontaneous polarization of up to 
$100\,\mu\rm{C/cm}^{2}$\cite{art_lebeugle_2007}.
The presence of multiple equivalent polarization states leads to the emergence of ferroelectric domains that can be separated by 180$^\circ$, 109$^\circ$, or 71$^\circ$ domain walls (indicating the angle between polarization vectors in adjacent domains)\cite{Catalan2009}.
%
%
The domain structures in BiFeO$_3$ can be quite complex, depending on the sample's shape, history, and boundary conditions. Most of the domain imaging has been done on thin films~\cite{art_zavaliche_2006,art_daumont_2010}; owing to the problematic synthesis of high-quality samples~\cite{art_lebeugle_2007} single-crystal studies are rather rare. 

Notwithstanding, one of the most unusual and exciting domain structures can be observed in single crystals grown (from flux) at temperatures just below the Curie temperature $T_{\rm C}$~\cite{art_berger_2012}. 
%
As exemplified in the transmission electron microscopy (TEM) dark-field images in Fig.\,\ref{fig_TEM}, these crystals host a distinctive structure of alternating straight and zigzag domain walls.
%
This extremely high density of self-similar triangular nanoscopic domains has relatively well-defined dimensions.
As can be seen in Fig.~\ref{fig_TEM}, the zigzag patterns exhibits different inclinations with respect to the straight interfaces depending upon the orientation of the crystal in TEM.
%
When viewed in projection (Fig.~\ref{fig_TEM}e), the regular zigzag sometimes breaks into an array of triangles, which are difficult to distinguish from each other.

It took several studies~\cite{art_berger_2012,art_jia_2015,art_condurache_2023} 
before the geometry of the observed domain structure was convincingly explained as being composed solely of 180$^\circ$ domains 
\cite{art_ge_2023_AFM, art_ge_2023_MIC}.
The sharp, straight interfaces, separated from each other by 50-100$\,{\rm nm}$, were identified as atomically thin non-stoichiometric layers with a preferential $(112)$ orientation formed during crystal growth, containing edge-sharing FeO$_6$ octahedra.\footnote{Throughout this article, for simplicity we use pseudo-cubic indexing of the BiFeO$_3$ unit cell, taking the spontaneous polarisation vector to lie along $[111]$.}
The non-stoichiometry leads to a negative charge density at the interface of approximately -68\,$\mu$C/cm$^2$ to -110$\mu$C/cm$^2$ \cite{art_maclaren_2013, art_maclaren_2014, art_li_2017}. 
Therefore, a complete ferroelectric-polarization reversal takes place at the thin interface, and a 180$^\circ$ head-to-head domain wall is formed (that is, ferroelectric polarization in adjacent domains points towards each other)~\cite{art_ge_2023_MIC}.
%

Consequently, there must be a second 180$^\circ$ domain wall between the straight interfaces
and this is the one that forms a zigzag pattern.
The zigzag was recently identified to be a projection of a relatively regular landscape of triangular pyramids with apices pointing along the direction of spontaneous polarization~\cite{art_ge_2023_AFM}. In other words, rows of pyramidal domains aligning in specific directions give the specific saw-tooth contrast for the 180$^\circ$ charged (tail-to-tail) domain wall seen in TEM.
However, while the geometry of the peculiar pattern was determined, its physical origin remains unclear.

The charged (head-to-head/tail-to-tail) domain walls always need to be compensated; otherwise, they cannot exist for electrostatic reasons. 
The bulk sample was found to be perfectly insulating, indicating the absence of mobile carriers or macroscopically contiguous conducting domain walls. Therefore, the the compensation of the zigzag wall is likely ionic. It is nevertheless difficult to foresee any reason
why the ionic compensation should lead to such a complicated domain pattern, with a relatively large surface area in comparison with a planar wall.
%
Another stunning feature of this domain structure is its stability: it does not show any response to sample manipulation, such as cutting or polishing (the samples also remained intact for many years of storage). This is extremely unusual, in particular for such a large polarization magnitude. Also, the response to an applied electric field is minor~\cite{art_condurache_2023}, indicating exceptional stability of the structure. 

Domain walls with a zigzag structure are found in ferroelectric materials.\cite{art_han_2014, art_denneulin_2022, Ricote} They have been addressed in
several theoretical studies,\cite{Misirlioglu2012, Cheah, art_zhang_2020, art_atkinson_2022, art_cornell_2023} which include various assumptions about the spatial distribution of electrostatic charges. 
However, to the best of our knowledge, none of the proposed mechanisms directly applies to the three-dimensional pyramidal domain structures observed in bulk BiFeO$_3$ crystals discussed here.





Here, motivated by the recently developed theory of zigzag domain patterns in ferroelectric materials \cite{art_marton_2023}, we apply the phase-field method
within the Ginzburg-Landau-Devonshire model to provide a theoretical foundation for the appearance of periodic pyramidal 180$^\circ$ domain walls in BiFeO$_3$. We show that a simple assumption/model/architecture of negatively charged thin planes alternating with positively charged thick layers
produces triangular pyramids whose periodicity, dimensions and arrangement are in excellent agreement with experimental observations.  We discuss how the properties of BiFeO$_3$ produce this particular structure in comparison with other materials, using as an example the prototypical ferroelectric BaTiO$_3$.



\section{Results}\label{sec_results}


\begin{figure}[h]
\centering
\includegraphics[width=1.00\textwidth]{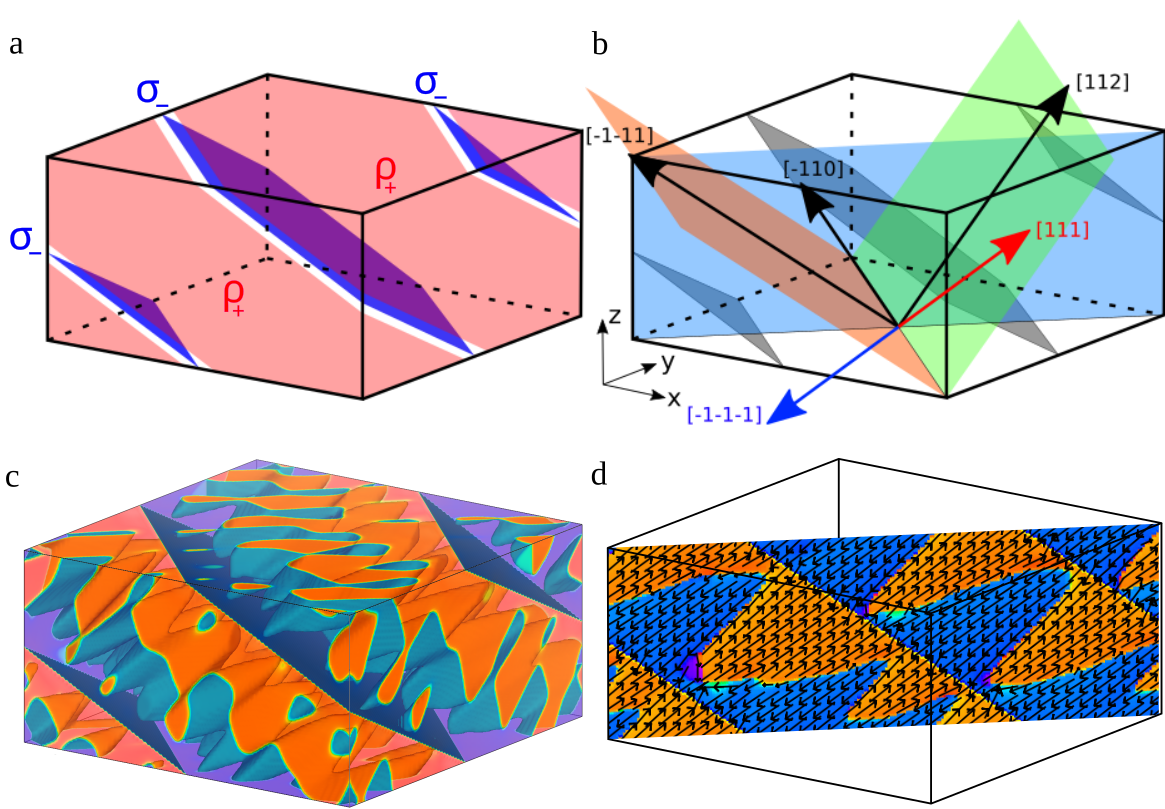}
\caption{The supercell utilized in phase-field simulations.
a) A negative charge planar density $\sigma_{-}$ (blue) is localized within the $(112)$ straight interfaces, and the positive charge density $\rho_{-}$ (light red) is diluted in the volumes between them. A small region close to the $(112)$ interfaces is left electrically neutral.
b) important directions and planes with respect to the supercell. The $(112)$-oriented negatively charged straight interfaces are in gray.
c) Domain structure emerging inside the supercell. Clearly visible are straight interfaces and the domain wall between them defined by ${\mathbf P=0}$. The color indicates the direction of ferroelectric polarization: the red region has polarization close to $[111]$, in the blue region it is opposite. 
d) Domain structure visualized on the $[\bar{11}0]$ cross-section of the simulation box, \REV{displayed using the HSL color space, with domains polarized along $[111]$ and $[\bar{1}\bar{1}\bar{1}]$ shown in orange and blue, respectively. At the domain wall, where polarization deviates from these primary directions, additional colors are visible.} The local polarization is indicated by small arrows. 
}
\label{fig_supercell_and_directions_scheme}
\end{figure}


%
Our simulations feature a regular array of negatively charged straight layers
with a planar charge density of $\sigma_{-}$, 
corresponding to a complete ferroelectric-polarization reversal (head-to-head).
We know from experiment \cite{art_ge_2023_MIC} that the orientation of the straight layers and thus the system's stacking direction is $[112]$. This orientation requires an appropriate configuration of the simulation box, which is visualized in Fig.\,\ref{fig_supercell_and_directions_scheme}a.
Between the negatively charged straight interfaces, we include a positively charged layer, 
with a charge volume density $\rho_{+}$ that exactly balances the negative charge in the straight interfaces. Such charge can originate, for example, from oxygen vacancies, a common positively charged ionic defect in perovskite ferroelectrics \REV{(the density of such defects is estimated to be $\approx10^{20}$\,vacancies/cm$^3$ in Sec.\,\ref{sec_methods_phase-field})}.
The charges $\rho_{+}$ and $\sigma_{-}$ interact with the polarization through their electrostatic field.
%


The simulations start as very small, randomly generated, ferroelectric-polarization vectors.
%
Immediately after the beginning of the optimization process, a head-to-head domain wall appears at the straight interface, and an irregular tail-to-tail wall forms within the charged layer, spanning its entire thickness.
As the simulation progresses, the domain structure becomes coarser, 
peaks reduce in number, and they develop into distinctive pyramids (Fig.\,\ref{fig_supercell_and_directions_scheme}c, d) with triangular bases. 

\begin{figure}[h]
\centering
\includegraphics[width=1.00\textwidth]{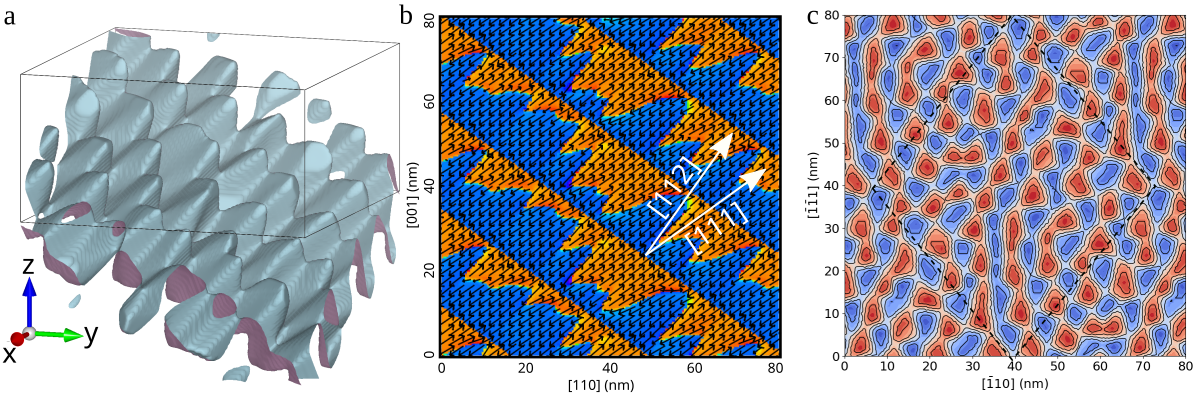}
\caption{
Simulation of domain patterns with a spacing of 16\,nm between straight interfaces.
a) VA pyramidal-array tail-to-tail domain wall running through a positively charged layer. The wall is defined as an isosurface where the polarization projected to the $[111]$ direction is zero, and it is unwrapped from the shown simulation box for better presentation.
%
b) Polarization projected onto a $(\bar{1}10)$ plane.
c) The distance of the wall from the center of the positively charged $[112]$-layer measured along the $[111]$ direction. Note the triangular pyramid's cross-section.
The planar plots throughout this paper cover an area of $80\times 80$\,nm$^2$.
Therefore, the domain-structure motifs are repeated with the appropriate periodicity of the simulation box (the irreducible part of the plot is marked with a dashed line in panel c).
} 
\label{fig_first_figure}
\end{figure}

The final results of such simulations are presented in Fig.\,\ref{fig_first_figure}.
The three-dimensional nature of the tail-to-tail domain wall is depicted in Fig.\,\ref{fig_first_figure}a, consisting of an array of pyramidal spikes (straight walls are omitted for clarity).
Fig.\,\ref{fig_first_figure}b illustrates the polarization projected onto and displayed on the plane $(\bar{1}10)$, which is the same as the projection plane in TEM Figures~\ref{fig_TEM}d-f. This visualization captures both the straight (head-to-head) and zigzag (tail-to-tail) 180$^\circ$ domain walls. It shows the re-entrant \REV{(`overhanging' facets, with an orientation more than $90^\circ$ away from the average domain wall plane) nature of the domain wall when inspected along $[\bar{1}10]$,} as reported in Ref.\,\cite{art_ge_2023_MIC}. Finally, it shows that the local polarization directions predicted by simulations significantly deviate from the pyramids' axes (we define the axis of a pyramid as a line running along the direction $[111]$ through the pyramid's apex). It is important to note that the cut plane in Fig.\,\ref{fig_first_figure}b does not, in most cases, go directly through the pyramid's apex, potentially creating the false impression that the pyramids terminate within the layer (which is never the case as proved by the Fig.\,\ref{fig_first_figure}a).
The structure of the tail-to-tail domain wall in the plane of the charged layer is visualized in Fig.\,\ref{fig_first_figure}c. We plot here distance of the domain wall from the center of the charged layer, red and blue colors mark the wall which is above and below the center of the layer, respectively (description of the distance measurement is in the Methods section).

 %

\begin{figure}[h]
\centering
\includegraphics[width=1.0\textwidth]{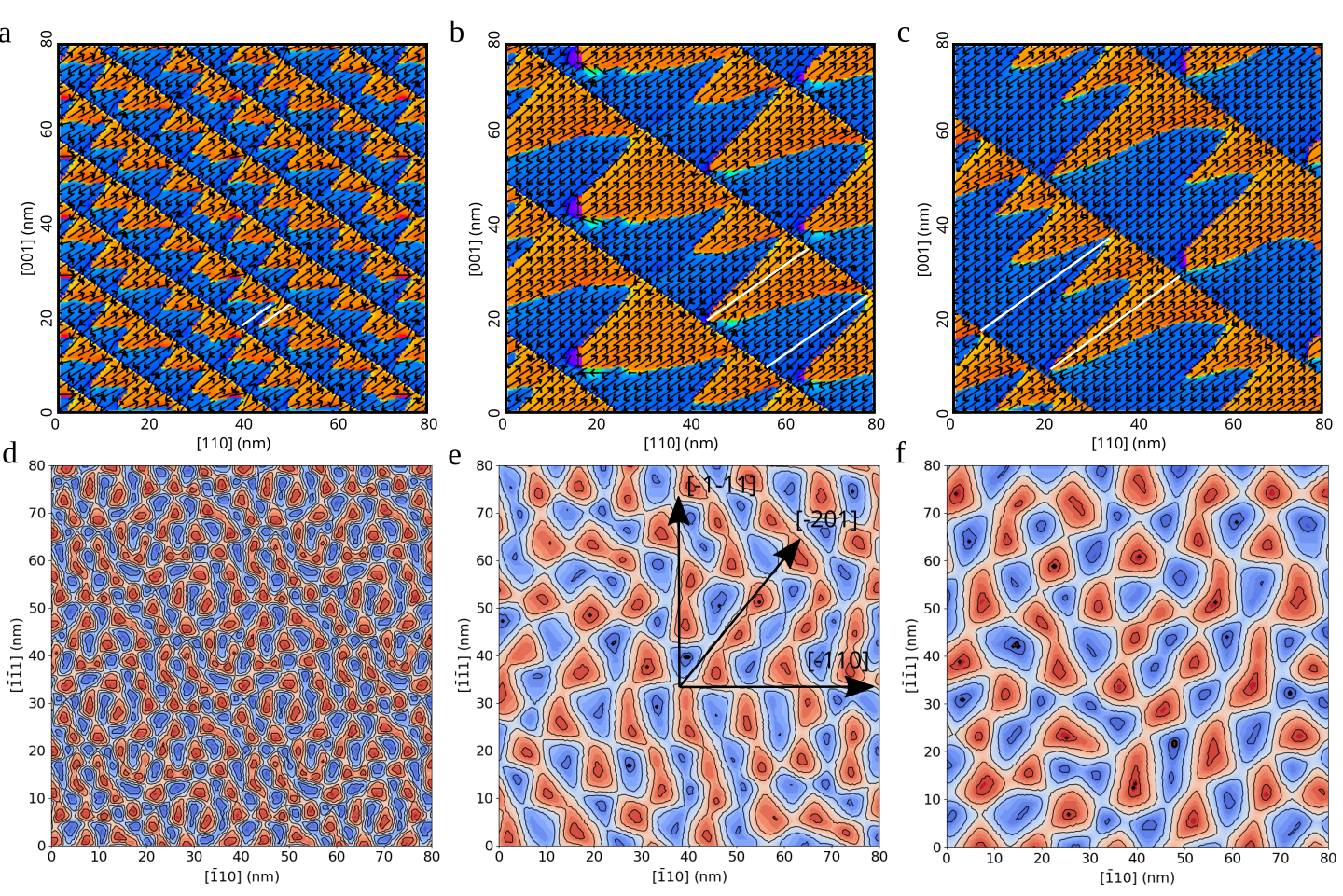}
\caption{The results of simulations with distances of straight interfaces $w_{\rm straight}$ being 8\,nm (a, d), 24\,nm (b, e), and 32\,nm (c, f). Axes of selected pyramids are indicated. In panel e, we show directions along which we identify the formation of chains of pyramids.}
\label{fig_cuts_color_different_sizes}
\end{figure}
Fig.\,\ref{fig_cuts_color_different_sizes} summarizes the final domain structures from simulations with distances of straight layers $w_{\rm straight}$ of 8, 24, and 32 nanometers.
For large range of $w_{\rm straight}$, the obtained domain patterns are very similar.
%
The emerging domains tend to have triangular cross-sections in the $(112)$ plane and form chains along directions close to $[\bar{1}\bar{1}1]$, $[\bar{1}10]$, and $[\bar{2}01]$.


\section{Discussion}\label{sec_discussion}

\begin{figure}[h]
\centering
\includegraphics[width=0.99\textwidth]{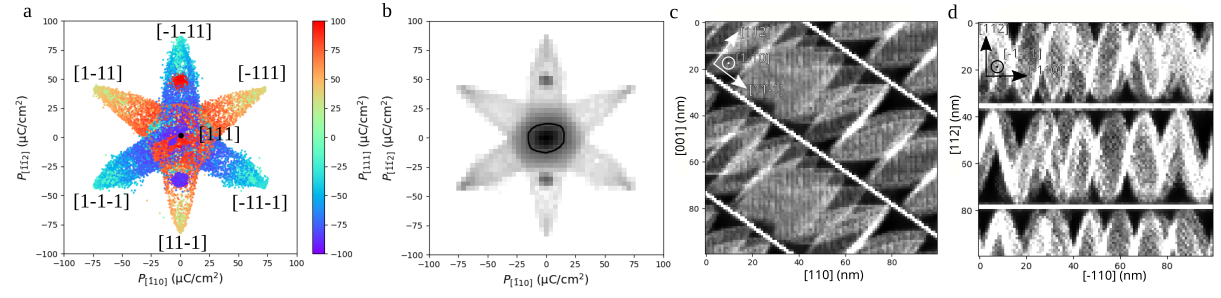}
\caption{
a) Distribution of transverse part of ferroelectric polarization for the complete simulation volume. The magnitude of longitudinal component of polarization (along $[111]$) is expressed by the color. There are preferred directions along which the polarization deviates from the ground state.
%
b) Histogram (in logarithmic scale) showing the number of counts of the transverse polarizations plotted in panel a. The contour is set to enclose the area with more than 80\% of all polarization arrows in the simulation box showing that, by far, most of the polarization vectors are aligned along the $[111]$ direction or very close to it. Large transversal polarization vectors (the rays of the star) strongly prefer to be aligned along the directions of
spontaneous states available in BiFeO$_3$, i.e. $\left<111\right>$ directions.
c) The domain structure as visualized in transmission along the probing direction of $[1\bar{1}0]$ revealing asymetric triangles, and 
d) along the direction $[\bar{1}\bar{1}1]$ showing symmetric triangles.
}
\label{fig_visualizationTEM_perpendicularP}
\end{figure}

The domain structures obtained in phase-field simulations are in excellent agreement with the experimental observations in many respects.
First, the simulated pyramids have a triangular cross-section and thus they always appear triangular,
independent of the viewing direction, in accordance with the analysis in Ref.\,\cite{art_ge_2023_AFM}. Moreover, the pyramids have varying outlines when viewed from different directions: \REV{they appear symmetric (see Fig.\,\ref{fig_TEM}a-c) when viewed along $[\bar{1}\bar{1}1]$ and asymmetric (see Fig.\,\ref{fig_TEM}d-f) when viewed along $[\bar{1}10]$} (both these directions are perpendicular to the system's stacking direction of $[112]$ and when viewed along these directions, straight interfaces are edge-on and appear as lines).
In order to quantify this aspect, we attempt to visualize how our phase-field-simulated domain walls might be seen in the transmission experiment along the directions $[1\bar{1}0]$ and $[\bar{1}\bar{1}1]$ in Fig.\,\ref{fig_visualizationTEM_perpendicularP}c and d, respectively. Although the pyramids obtained in simulations are not aligned along the probing direction, this visualization produces contrast resembling the TEM images. In particular it reveals asymmetric and symmetric triangles, in agreement with the experimental TEM data.
This is in accordance with the TEM micrographs shown in Fig.\,\ref{fig_TEM}. Next, while pyramids have re-entrant facets, the angle of these facets is not sufficient to reach the the $[111]$ polar axis, i.e. all domain walls are found at an increasing distance from a $[111]$ vector that passes through the apex of a pyramid, in accordance with the simulations.
Moreover, the simulations reveal the tendency for the pyramids to order into chains and arrays is in agreement with the TEM observation in Fig.\,\ref{fig_TEM}e, where an array of regularly arranged pyramids is clearly visible.
In addition, the simulated pyramids are rather narrow, in accordance with experiment.
And finally, the pyramid's height seems to correlate with the thickness of the layers.

Previous theoretical works have suggested a link between positive space charge and head-to-head zigzag domain structures~\cite{Misirlioglu2012,Cheah,art_marton_2023}.
Recently, a system of negatively charged thin interfaces
and oppositely charged thick layers was studied theoretically in tetragonal ferroelectric PbTiO$_3$\cite{art_marton_2023}. 
%
%
It was shown that such alternating system of charges implies presence of a straight head-to-head wall at the negatively charged plane and a tail-to-tail wall inside the positively charged layer. The tail-to-tail wall was shown to naturally form a rooftop-like zigzag pattern if the charged layer is sufficiently thick. Such ferroelectric-polarization arrangement was shown to allow for compensation of the positive defect charges located inside the layer through a favorable spatial variation of ferroelectric polarization inside individual triangles of the zigzag: the rotation of polarization vectors gives rise to a polarization bound charge according to the formula $\rho_{\rm \mathbf P}=-{\rm div}{\mathbf P}$, which approximately compensates
the positive charge of the layer, thus preventing the formation of energetically costly electric fields produced by uncompensated charges.


This compensation mechanism has three important hallmarks. First, the zigzag pattern covers the entire thickness of the charged layer. Second, the walls form straight planar segments, i.e. they appear linear on a cut. 
%
%
Finally, the zigzag pattern has a natural periodicity, which arises from the balance of two energy contributions: i) the excess energy of polarization rotation away from the energetically most favorable spontaneous state and ii) the energy elevation due to the increased surface energy of the wall.
The natural periodicity of the zigzag wall $W_\mathrm{natural}$ was shown\cite{art_marton_2023} to depend on the thickness of the charged layer $w_{\rm straight}$ and the material's properties, such as transverse permittivity $\chi_\perp$, the magnitude of spontaneous polarization $P_\mathrm{s}$, and surface wall energy density $\mu$.
The derivation of a similar formula for the pyramidal wall would be much more challenging and exceeds the scope of this paper. From the nature of the presented situations it nevertheless follows that such a natural size of the pyramids exists, which explains why the triangular domains observed using TEM are so regular/self-similar. Notice that in Fig.\,\ref{fig_TEM}b, in which the pyramids have very different heights, these pyramids were formed within deformed "straight" layers of varying thicknesses, in accordance with these insights.\footnote{We have attempted to manifest the dependence of the pyramid's shapes from TEM on the distance of straight walls, but the results were inconclusive. This was partly due to very similar distances of the straight layers in most cases, partly because the apices of the pyramids could not be resolved with high enough accuracy.}

%
%

%
The domain structure obtained from simulations for the BiFeO$_3$ is rooted in the same compensation mechanism. Therefore the pyramids go through the entire thickness of the charged layer,
%
have approximately planar faces, and the pyramidal system converges to self-similar shapes. These aspects predicted by simulations are also in agreement with the TEM experimental observations, as discussed above and seen in Fig.\,\ref{fig_TEM}. In accordance with the proposed mechanism, the simulations predict that the farther from the pyramid's axis, the greater the deviation of $\mathbf{P}$ from the direction $[111]$. This rotation yields the bound charge $\rho_{\rm \mathbf P}=-{\rm div}\mathbf{P}$ that compensates for the defect charge $\rho_{+}$.


The idea that the formation of the pyramidal tail-to-tail domain wall is caused by the presence of a positive space (point-defect) charge distributed rather homogeneously inside the layer brings us back to one important aspect of the TEM data. A closer inspection of the TEM images reveals that the pyramids actually do not span the whole width of the layer, even in thick parts of the specimen where they are fully captured in the lamella.  This can be caused by the depletion of positively-charged point defects close to the straight interface.
Close to the edge of the wedge-shaped specimen for the TEM observation (Fig.\,\ref{fig_TEM}c,f), the reason is geometric: the very thin specimen only contains a section through the pyramids, and the (uncontrolled) position of this section does not usually cut through their peaks.

%
The most significant 
feature of the conducted simulation is formation of triangular-base pyramids, in excellent accord with the observations using HRTEM and interpretation in Ref.\,\cite{art_ge_2023_AFM}. The reason is the three-fold symmetry of the energy surface \REV{normal to} the $[111]$-direction. 

In principle, any deviation of the ferroelectric polarization from $[111]$ could lead to a favorable bound charge $\rho_{\rm \mathbf P}$, but some directions are preferential because they are energetically cheaper. In order to manifest this, we show the compilation of all ferroelectric-polarization vectors (irrespective of their location in the simulation box) projected on $(111)$ in Fig.\,\ref{fig_visualizationTEM_perpendicularP}a. Clearly, the polarization with large magnitude tends to deviate from $[111]$ dominantly along six directions. Three of them are active when the polarization points up, along $[111]$, and the other three when the polarization points down, along $[\bar{1}\bar{1}\bar{1}]$.
Notice that most of the polarization vectors are aligned along the $[111]$ or $[\bar{1}\bar{1}\bar{1}]$ directions and only a minority of polarization vectors are very far from these, as evidenced in  Fig.\,\ref{fig_visualizationTEM_perpendicularP}b.

\begin{figure}[h]
\centering
\includegraphics[width=0.99\textwidth]{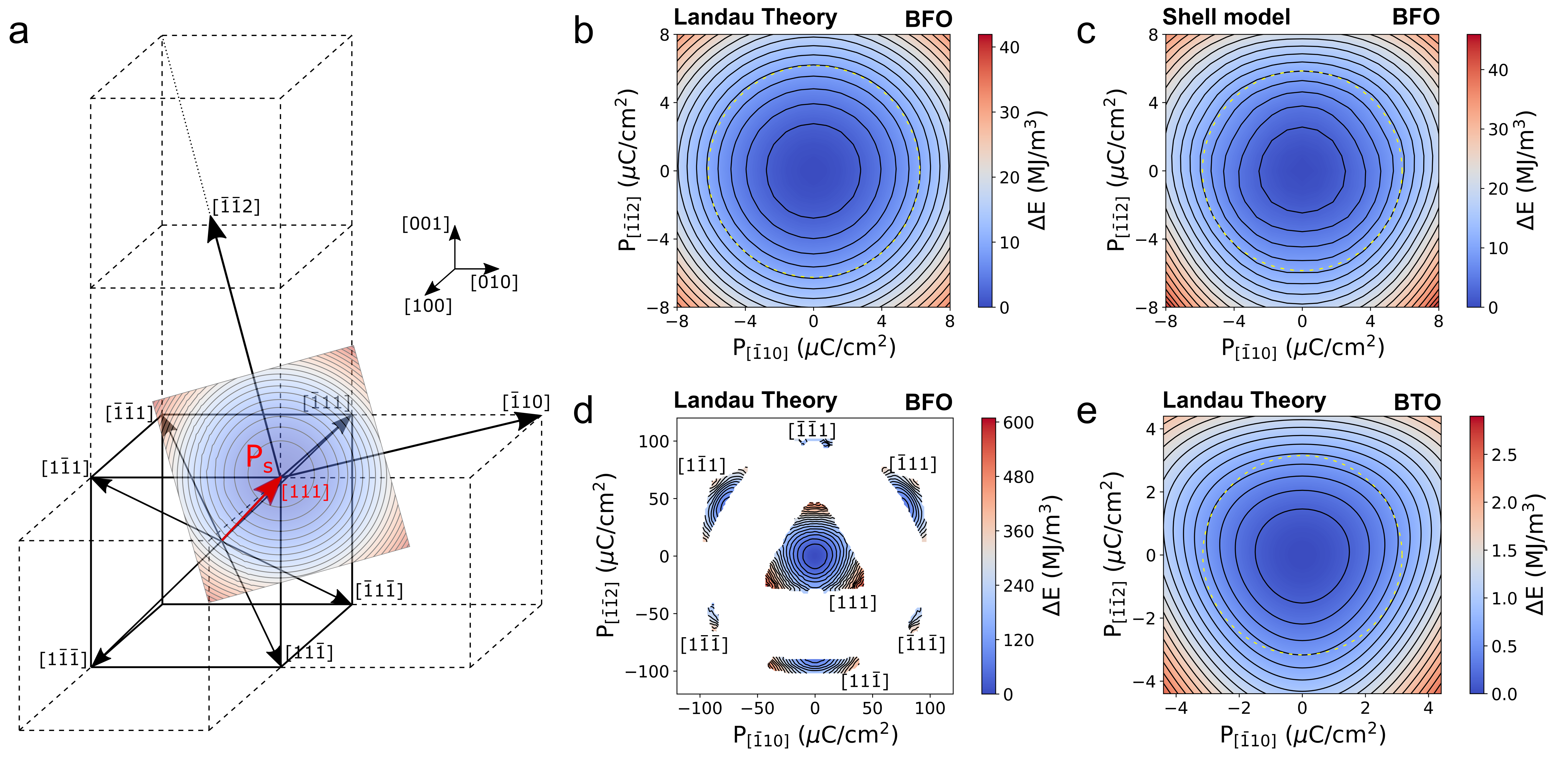}
\caption{Anisotropy of the energy surface for transverse part of the ferroelectric polarization $\Delta\mathbf{P}$ ($\Delta\mathbf{P}\perp\mathbf{P}_{\rm s}$) as probed by application of electric field perpendicular to $\mathbf{P}_{\rm s} \parallel [111]$, and evaluation of the resulting ferroelectric polarization and its energy. In calculations and simulations, the materials's strain was kept fixed and equal to the spontaneous strain, which is the relevant mechanical state in numerical simulations presented in this work.
a) Sketch of the orientation of the energy surface with respect to the coordinate system and spontaneous directions in BiFeO$_3$.
b) Energy map evaluated using the GLD model for BiFeO$_3$\cite{art_marton_2017} utilized here in simulations.
c) Energy map obtained using independent shell-model molecular dynamics simulations for BiFeO$_3$. Excellent agreement with b is obtained; in particular, both plots indicate that the electrically soft directions are along $[11\bar{2}]$ and the other two directions that are related by the three-fold rotation along $[111]$.
Both b and c feature a relatively small range of polarization values (8\,$\mu$C/cm$^2$ is about 10\% of the spontaneous value), which are, according to figure \ref{fig_visualizationTEM_perpendicularP}b, the most important range of values.
d) A large-polarization energy surface for BiFeO$_3$, showing discontinuity upon 71$^\circ$ and 109$^\circ$ switching from the $[111]$ state for large values of the electric field.
e) The small-polarization energy map for GLD model of BaTiO$_3$ as parametrized in Ref.\,\cite{art_bell_2001}.
}
\label{fig_landau_potential_perp_to_111}
\end{figure}

In order to understand the origin of this anisotropy, we inspect the dependence of the potential energy for the transverse displacement of the polarization vector from the spontaneous state along $[111]$ in Fig.\,\ref{fig_landau_potential_perp_to_111}b. The plots are obtained for a single unit cell with polarization deviation driven by an applied external electric field. Considering 
importance of this energy surface for this work, we also evaluate the corresponding potential map using a more accurate atomistic shell-model in Fig.\,\ref{fig_landau_potential_perp_to_111}c. These two independent simulations exhibit an almost perfect agreement and give us confidence that the physics predicted by the GLD model is correct. The predicted potential limited to small values of ferroelectric polarization is only weakly anisotropic.
%
Because the deviation of polarization from the $[111]$ direction obtained in simulations is rather large, we also show a polarization map for a larger range of transversal polarizations, for which the anisotropy is well visible. In this figure, in contrast to the domain-structure simulations, where the polarizations vary  smoothly (confined by the defect charge), some parts of the energy surface are not accessible due to field-driven polarization switching, which leads to sudden jumps.

The obtained energy surfaces in Fig.\,\ref{fig_landau_potential_perp_to_111}b,c,d are anisotropic, and consequently, along some directions \REV{the energetic cost of polarization deviation $\Delta\mathbf{P}$ is smaller}. It is no surprise that these favorable directions are along $\left<111\right>$, i.e. symmetrically equivalent spontaneous states present in BiFeO$_3$.
Although the anisotropy of the energy dependence on ferroelectric polarization is rather small (owing to the mechanical clamping), it manifests itself strongly in the simulations and is the reason for the formation of domains with a triangular cross-section
and not a zigzag wave-like pattern like in PbTiO$_3$, where the energy surface has a four-fold symmetry.
%
\footnote{The discussion here focused solely on the anisotropy of Landau-energy properties. It should be pointed out that the shape and orientation of the pyramids will also be to some extent affected by the gradient term in the GLD model, which is indeed three-fold symmetric with respect to the $[111]$ axis as well.}

%
Another interesting feature observed in simulations is the tendency of the pyramids to arrange in chains (Figs.\,\ref{fig_first_figure}c and \ref{fig_cuts_color_different_sizes}d,e,f),
especially in the directions close to $[\bar{1}\bar{1}1]$, $[\bar{1}10]$, and $[\bar{2}01]$. Such chains are consistent with TEM observations, where the pyramids seem to appear ordered along some directions. The formation of the chains can be seen as a compromise between a pyramidal structure, obeying the  three-fold Landau anisotropy and a rooftop-like structure, which has a smaller overall domain wall surface.




In this context, we found it interesting to check whether zigzag rooftop-like walls represent a viable solution for BiFeO$_3$. We tested this by conducting phase-field simulations started from a wave-like polarization profile with wave-vectors $\mathbf{k}\parallel[\bar{1}10]$ and $\mathbf{k}\parallel[\bar{1}\bar{1}1]$ (both propagation vectors perpendicular to $[112]$).
%
The rooftop-like pattern was found metastable for the $\mathbf{k}\parallel[\bar{1}10]$ with the energy of the system slightly higher compared to pyramidal domain walls. The $\mathbf{k}\parallel[\bar{1}\bar{1}1]$ is unstable and decomposes into pyramids. It is possible, that in real material,  rooftop-like solutions might be stable, allowing for appearance perfect triangles along some directions.

The proposed scenario of the distribution of positive ionic defects in the region between negatively charged reconstructed head-to-head domain walls is adopted here as an assumption for the simulations. This assumption not only explains the formation of pyramidal domain-wall patterns,  as shown by the simulations, but it also accounts for the remarkable stability of the zigzag domain pattern.

For preparing wedge-shaped samples for TEM, the BiFeO$_3$ crystals were cut using a focused ion beam technique. Given the high spontaneous polarization in BiFeO$_3$, one might expect that surface charges produced by the ferroelectric polarizations pointing out of the cut surface would lead to strong depolarizing fields, causing significant rearrangement in the domain structure. However, the domain structure remained intact even in the thinnest parts of the specimens. Furthermore, the domain structure has shown long-term stability (over 10 years since crystal growth) and has also been observed to remain rather stable under the application of an external electric field \cite{art_condurache_2023}.

This stability is in line with the concept of ferroelectric polarization compensating for defect charges: the polarization-bound charge is locally balanced by charged defects and vice versa, ensuring no uncompensated polarization appears at the surface when the crystal is cut. Additionally, the domain structure cannot easily shift under an applied electric field, as it compensates the charge. It is likely that some or many defects might have migrated toward regions near the pyramidal walls despite their limited mobility, driven by minor electric imbalances or mechanical factors; the defects in ferroelectrics are known to segregate at domain walls. These migrated defects can further pin the domain walls, enhancing the stability of the domain pattern.

%
The zigzag wall and the presence of the charged layer are closely linked; for electrostatic reasons, it is difficult to envision one without the other, indicating that they are likely to form simultaneously.

\begin{figure}[h]
\centering
\includegraphics[width=1.0\textwidth]{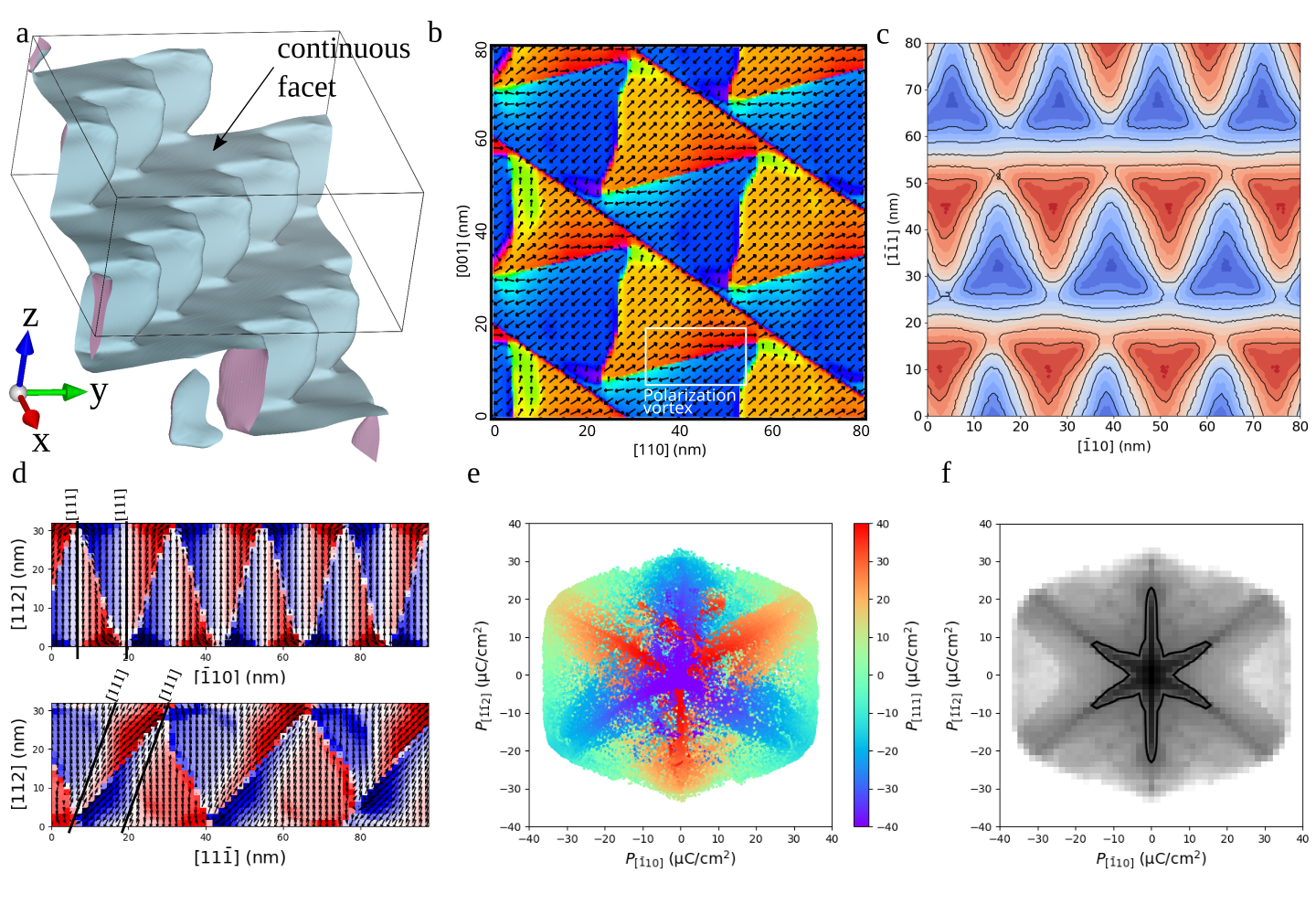}\\
\caption{
Phase-field simulations for the rhombohedral BaTiO$_3$ with the distance of straight interfaces of 32\,nm.
a) 3D visualization of the wall surface with the indication of the continuous facet.
b) Ferroelectric polarization visualized in the plane $(\bar{1}10)$.
c) Distance of the domain wall from the center of the charged layer measured along $[111]$.
\REV{d) Deviation of the polarization from the pyramid's axis. Polarization is projected to two selected planes, color indicates polarization component perpendicular to [111] direction. Red: positive, blue: negative, white: no deviation.}
e) Distribution of the polarization vector in the plane $(111)$, as collected from the entire simulation box, 
f) Corresponding log-histogram with the contour enclosing 80\% of polarization arrows obtained in the simulation.
}\label{fig_BTO_cuts}
\end{figure}
%
%
To illustrate how the mechanism of polarization-originated compensation for defect charge within a thick layer might manifest itself in other ferroelectrics, we investigated a prototypical ferroelectric, BaTiO$_3$, at low temperatures in the rhombohedral phase. Results obtained under conditions similar to those used for BiFeO$_3$ are shown in Fig.\,\ref{fig_BTO_cuts} for a domain wall spacing of $w_{\rm straight}=32$\,nm.

We predict that a pyramidal domain structure develops inside a charged layer in BaTiO$_3$ as well. The main distinction from BiFeO$_3$ is a strong tendency for the pyramids in BaTiO$_3$ to form chains aligned along $[\bar{1}10]$ direction, accompanied by the formation of a continuous facet (see Fig.\,\ref{fig_BTO_cuts}a), 
The pyramids in BaTiO$_3$ are predicted to be much broader than those of BiFeO$_3$.
The difference can be rationalized by the fact that BaTiO$_3$ dielectrically softer, i.e. the potential energy for transverse polarization in BaTiO$_3$ is shallower (and more isotropic), as visualized in Fig.\,\ref{fig_landau_potential_perp_to_111}e. This permits greater deviation of ferroelectric polarization from the spontaneous direction, resulting in wider pyramids and consequent reduction of domain-wall surface energy.

In Fig.\,\ref{fig_BTO_cuts}b, we present the spatial distribution of ferroelectric polarization and its gradual deviation from the spontaneous directions. Notably, a polarization vortex appears in the wall region, resembling the one identified in Fig.\,6e of Ref.\,\cite{art_ge_2023_AFM}. Due to their larger base, the pyramids in BaTiO$_3$ are not re-entrant. 

The distance of the domain wall from the center of the positively charged layer is shown in Fig.\,\ref{fig_BTO_cuts}c. In this representation, the continuous facet resembles the chain of pyramids observed in BiFeO$_3$ in Figs.\,\ref{fig_cuts_color_different_sizes}e,f,g (where the chains nevertheless appear along multiple directions). This suggests that a similar continuous facet may also be present in BiFeO$_3$, but its energy advantage may be too small for the phase-field optimization to pick this more ordered state over the irregular arrangement of Fig.~\ref{fig_cuts_color_different_sizes}, especially with the randomized initial conditions.

\REV{In Fig.\,\ref{fig_BTO_cuts}e, we illustrate one of the most important aspects of the pyramidal domain wall: the gradual deviation of ferroelectric polarization from the spontaneous direction [111] as one moves away from the pyramid axis. The polarization (arrows) is projected onto two mutually perpendicular planes, both of them containing the normal to the straight interface [112]. The cross-sections were deliberately chosen to pass through the apexes of the pyramids. The color in the inset represents the polarization component perpendicular to the spontaneous [111] direction, ranging from blue for negative values, through white (where the polarization has no perpendicular component), to red for positive deviations. It can be seen that the pyramid axis always lies in the white region, with the color intensity increasing as one moves away from the axis.}

\REV{Finally, the hexagon-shaped distribution of ferroelectric polarization perpendicular to the spontaneous direction, shown in Figs.\,\ref{fig_BTO_cuts}e differs considerably from the star-like shape obtained for  BiFeO$_3$ in Fig.\,\ref{fig_visualizationTEM_perpendicularP}a.} This highlights the importance of local polarization anisotropy for the shape and connectivity of pyramidal domains within the charged layer.

The results for BaTiO$_3$ demonstrate the proposed mechanism’s broad applicability across various perovskites and underscore the versatility of the resulting domain configurations, which depend on each material’s intrinsic properties. It is worth noticing that a zigzag pattern was recently observed experimentally in a tail-to-tail charged wall within a tetragonal BaTiO$_3$ thin film, grown using the pulsed laser deposition (PLD) technique on a $[110]$-oriented orthorhombic NdScO$_3$ substrate.\cite{art_denneulin_2022}


\section{Conclusions}\label{sec_Conclusion}

Phase-field simulations of the domain structures in BiFeO$_3$ reveal realistic pyramidal patterns that align closely with experimental observations, addressing a long-standing puzzle in ferroelectric research. We propose a theory in which 
these patterns form to compensate for defect charges, which arise within the material during growth under specific conditions. This charge compensation is achieved through favorable variations in ferroelectric polarization within the pyramidal domain pattern. Our conclusion is strongly supported by the high correlation between our simulations and experimental observations using TEM.

A key feature of the emerging domain structure is the regularity of the pyramidal pattern. The geometry of these pyramids result from a balance between the surface energy of the domain walls and the energy associated with ferroelectric polarization rotating away from its spontaneous direction. Thus, pyramid dimensions vary according to the specific material and structural parameters of the layered configuration.

These periodic, self-assembled domain arrangements could have significant implications in fields that require precise nanometric periodicity, such as optics. Additionally, the pyramidal domains may serve as regularly spaced carriers of topological defects, expanding the potential applications of ferroelectric and multiferroic materials. The insights provided in this paper are crucial for achieving controlled synthesis of high-quality pyramidal structures in perovskites for such targeted applications.

%
%
%
%
%

\section{Acknowledgment}\label{sec_acknowledgment}

We are grateful to Prof. Ana M. Sanchez and Dr. Wanbing Ge for fruitful discussions and some of the TEM images in Fig.~\ref{fig_TEM}. 

The research was supported by Czech Science Foundation Grant No. 24-11275S.
Computational resources were provided by the e-INFRA CZ project (ID:90254), supported by the Ministry of Education, Youth and Sports of the Czech Republic. 
This work was supported by the European Union’s Horizon 2020 research and innovation programme under grant agreement no. 964931 (TSAR) and by the Ferroic Multifunctionalities project, supported by the Ministry of Education, Youth, and Sports of the Czech Republic, Project No. CZ.02.01.01/00/22\underline{\,\,\,}008/0004591, co-funded by the European Union.

\section{Methods}\label{sec_methods}

\subsection{Sample preparation}
\label{sec_sample_preparation}

BiFeO$_3$ single crystals were grown by a generic flux grown method similar to that proposed by Kubel and Schmid.\cite{art_kubel_1993}
The grown flux was Bi$_2$O$_3$ mixed with B$_2$O$_3$ (4\% wt.) to reduce the melting point under 622$^\circ$C. The load was heat to 850$^\circ$C for 10h, then cooled to 610$^\circ$C with 0.6 K/h and finally to room temperature with 5K/min. BiFeO$_3$ (001)-oriented rosette-like pyramidal crystals, typically larger than 1 mm $\times$ 1 mm, with a 
thickness between 100 and 300 $\rm \mu m$, grown on the top of the melt were harvested by etching the flux with glacial acetic acid. The crystals were finally cleaned using 10\% HNO$_3$.

\subsection{TEM experimental details}
\label{sec_methods_TEM}
Specimens were prepared for TEM using a TESCAN Amber Ga$^+$ focused ion beam microscope operating at 30 kV (with 5 kV used for final polishing).  Lift-out sections were taken from the $(001)$ polished surface of a BiFeO$_3$ crystal, producing electron transparent lamellae with $[1\bar{1}0]$ and $[110]$ orientations.  A small wedge angle was used in the final thinning to obtain the thinnest possible section at the lamella edge.  Samples were examined in a JEOL 2100 LaB$_6$ TEM operating at 200 kV.

\subsection{Phase-field model and simulations of domain structures}
\label{sec_methods_phase-field}
The simulations of the domain structure were conducted using phase-field simulations in the framework of the Ginzburg-Landau-Devonshire model with the dissipative
Landau-Khalatnikov dynamics as implemented in the simulation package FERRODO2\cite{art_marton_2006, art_marton_2023}. 
The energy functional includes the Landau, gradient, elastic, electrostrictive and electrostatic contributions. 
Its form and parametrization for BiFeO$_3$ are taken from Ref.\,\cite{art_marton_2017} (and it is based on first-principles calculations). 
This parametrization leads to spontaneous values of $A_{\rm s}=14.36^\circ$,
$P_{\rm s}=91$\,$\mu$C/cm$^2$,
${e_{\rm s}}_{\rm xx}=0.0136$, and
${e_{\rm s}}_{\rm xy}=0.0012$ for polarization, G-ordered oxygen-octahedron tilts, diagonal and off-diagonal components of the strain tensor, respectively.
The gradient coefficients were chosen as $G_{11}=3.36879\times 10^{-11}$\,${\rm Jm^{3}C^{-2}}$, $G_{12}=-9.80011\times 10^{-11}$\,${\rm Jm^{3}C^{-2}}$, $G_{44}=1.72523\times 10^{-10}$\,${\rm Jm^{3}C^{-2}}$, for ferroelectric polarization and $G_{11}=1.18227\times 10^{-13}$\,${\rm Jm^{-1}deg.^{-2}}$, $G_{12}=-4.15832\times 10^{-13}$\,${\rm Jm^{-1}deg.^{-2}}$, $G_{44}=1.10073\times 10^{-12}$\,${\rm Jm^{-1}deg.^{-2}}$ for oxygen-octahedron tilt.
Discretization step $\Delta$=0.4\,nm is chosen similar to the lattice constant of BiFeO$_3$ in order to avoid pinning of walls by discretization\cite{art_marton_2018}. Previously, it has been shown that the carefully parametrized GLD model can be in good agreement with more accurate atomistic approaches\cite{art_taherinejad_2021, art_marton_2023}, for which our current target system is too large. 

The simulation box was constructed in order to host $(112)$-oriented negatively charged interfaces and the positively charged layer between them, see Fig.\,\ref{fig_supercell_and_directions_scheme}. Its shape is L$\times$L$\times$L/2. Considering the all-direction periodicity of the box, it contains a single $[112]$-oriented layer with the domain structure of interest. Distances $w_{\rm straight}$ of the $(112)$ interfaces considered  here are $\approx 8\,{\rm nm},\,\approx 16\,{\rm nm},\,\approx 24\,{\rm nm},\,\approx 32\,{\rm nm}$.

In simulations, the planar density of the charge on straight interface is chosen to be $\sigma_{-}=-2P_{\rm s}\times{\rm cos}(19.47)=-172$\,$\mu$C/cm$^{2}$, accounting for the angle of 19.47$^\circ$ between normal of the straight interfaces $[112]$ and direction of spontaneous polarization $[111]$. The positive charge has volume density of $\rho_{+}=-1\times\sigma_{-}/w_{\rm straight}$, where $w_{\rm straight}$ denotes the thickness of the charged layer. The interaction of ferroelectric polarization with the positive and negative charges are modeled through their electrostatic fields, which leads to the same result as explicitly considered charges, as discussed in Ref.\,\cite{art_marton_2023}. The field's magnitude increases linearly from one straight interface to the next, reaching zero at the center of the layer, then reverses abruptly at the straight interface, and increases again in the following layer. To prevent interaction between domain structures on the opposite sides of the straight interface, we introduce a $\sim$3~nm thick neutral layer between the positively and negatively charged regions, where the electrostatic field remains constant, as per Gauss's law.

\REV{The charge density $\rho_{+}$ can be used to evaluate a corresponding density of oxygen vacancies, which are considered a model positive defect providing the compensation of the charged zigzag wall. For the distance of straight interface of $50\,{\rm nm}$ (distance encountered in TEM observations) the density of oxygen vacancies of $1.07\times 10^{20}$ per cm$^3$, which means that approximately 0.23\,\% of oxygens in the perovskite system are missing. The density of oxygen vacancies is inversely proportional to the distance of straight interfaces, i.e. doubled distance between straight interfaces leads to half of the density.}

The simulations \REV{for BiFeO$_3$} were conducted in mechanically clamped conditions, the mechanical strain tensor is fixed locally to its spontaneous value throughout the simulation in accordance with the known non-ferroelastic character of the crystal containing only 180$^\circ$ domain walls\cite{art_ge_2023_MIC}. Both the ferroelectric polarization and the rotation of the oxygen octahedra are free to evolve. The initial condition for polarization is chosen either random or wave-like, the initial rotations of the oxygen octahedra attain spontaneous value. The zigzag pattern forms spontaneously within the layer, there is no other incentive needed\cite{art_marton_2023}. \REV{The oxygen-octahedron rotations follow the $a^-a^-a^-$ tilt system in Glazer’s notation~\cite{art_glazer_1972}, as expected for bulk BiFeO$_3$. Antiphase boundaries (APBs), i.e. a displacement of the oxygen octahedral tilt system by a single prototype unit cell, tended to coincide with the pyramidal wall when present. No APBs were observed experimentally. It seems that the effect of APBs on the emerging polarization domain structure is insignificant.}

Investigations on BaTiO$_3$ were conducted using Ginzburg-Landau-Devonshire with Landau and gradient potentials from Refs.\,\cite{art_bell_2001} and \cite{art_hlinka_2006}, resp. The complete set of utilized parameters is recorded as Model I in Table 4.6 of Ref.\,\cite{boo_ondrejkovic_2020}. The magnitude of the electric field and the spontaneous strain were modified in order to correspond to spontaneous values of BaTiO$_3$, the rest of the simulation setup remained unchanged. 

\begin{figure}[h]
\centering
\includegraphics[width=0.6\textwidth]{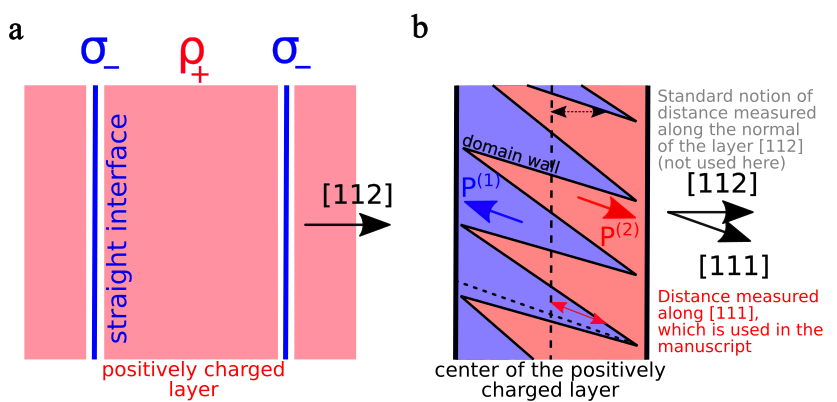}
\caption{
a) Schematics of the here considered distribution of electric charge. A negative charge located on a planar surface with a planar density of $\sigma_{-}$ regularly alternates with the positively charged region with a volume density of $\rho_{+}$.
b) Measurement of the distance of the domain wall (separating polarization blue domain ${\mathbf P}^{(1)}\parallel [\bar{1}\bar{1}\bar{1}]$ from the red domain ${\mathbf P}^{(2)} \parallel [111]$) from the center of the charged layer along the direction $[111]$ as utilized in Figures  \ref{fig_first_figure}c, \ref{fig_cuts_color_different_sizes}d,e,f and \ref{fig_BTO_cuts}c. Measurement along the stacking direction $[112]$ would lead to ambiguities, as shown by the black arrow for the overhanging domain.
}
\label{fig_schematics_of_distance_measurement}
\end{figure}

In this whole work, when we refer to the distance of the domain wall from the center of the charged layer, we mean the distance measured along the $[111]$ direction, as indicated in the Fig.\,\ref{fig_schematics_of_distance_measurement}. This choice allows us to avoid obtaining multiple distances when we measure perpendicular to the charged layer and the pyramids are re-entrant, and arrive to unambiguous result. 

In the paper we visualize the data obtained from simulations in such a way that they mimic the TEM experimental micrographs. We do not simulate here the acquisition of the TEM signal, but rather use the $\left|{\rm div} ({\mathbf P_{[111]}})\right|$ the absolute value of the divergence of the $[111]$-projected ferroelectric polarization vectors as an indicator for the presence of the charged domain wall in the material's bulk. This quantity is large on the charged wall and small in the domain interior. This quantity was integrated along the probing directions $[1\bar{1}0]$ and $[\bar{1}\bar{1}1]$,  and visualized in Fig.\,\ref{fig_visualizationTEM_perpendicularP}c and d. 

\subsection{Shell-model evaluation the energy surface}
\label{sec_methods_shell-model}

We have employed here an ab initio-based atomistic shell model potential for BiFeO$_3$ molecular dynamics simulations previously presented in Ref. \cite{ graf2014}. The model was previously successfully applied in calculations involving neutral domain walls in BiFeO$_3$~\cite{art_hlinka_2017}. Each atom is represented by core and shell, with its own position and charge. The potential energy includes pairwise short-range interactions as well as electrostatic Coulomb interactions. We have used supercells made of 14×14×14 perovskite unit cells (i.e., 13720 independent atoms) with periodic boundary conditions. Stable configurations under an external electric field were obtained by performing classical molecular dynamics simulations using DL POLY software \cite{dlpoly}. We use a time step of 0.2 ps and an equilibration time of 12 ps. After ensuring that the system is in equilibrium, we kept it evolving for an extra 4 ps, in which we recorded snapshots (every 0.05 ps) used to calculate the energy and polarization of the system.

\REV{\section{Data availability}\label{sec_data_code_availability}
The data that support the findings of this study and the computer codes are available from the corresponding author upon reasonable request.}

\bibliography{sn-bibliography}
\end{document}